# 3D Pseudo Stereo Visualization with Gpgpu Support

**Anas M. Al-Oraiqat**[1] and **Sergii A. Zori**[2]

[1]*Department of Cyber Security, Onaizah Colleges, College of Engineering & Information Technology، Onaizah, Kingdom of Saudi Arabia, P.O. Box 5371, and* [2]*Department of Computer Sciences and Technologies, SHEE «Donetsk National Technical University», Donetska Oblast, P.O.Box 85300, Ukraine*

anas_oraiqat@hotmail.com

*Abstract.* This article discusses the study of a computer system for creating 3D pseudo-stereo images and videos using hardware and software support for accelerating a synthesis process based on General Purpose Graphics Processing Unit (GPGPU) technology. Based on the general strategy of 3D pseudo-stereo synthesis previously proposed by the authors, Compute Unified Device Architect (CUDA) method considers the main implementation stages of 3D pseudo-stereo synthesis: (i) the practical implementation study; (ii) the synthesis characteristics for obtaining images; (iii) the video in Ultra-High Definition (UHD) 4K resolution using the Graphics Processing Unit (GPU). Respectively with these results of 4K content test on evaluation systems with a GPU the acceleration average of 60.6 and 6.9 times is obtained for images and videos. The research results show consistency with previously identified forecasts for processing 4K image frames. They are confirming the possibility of synthesizing 3D pseudo-stereo algorithms in real time using powerful support for modern Graphics Processing Unit/Graphics Processing Clusters (GPU/GPC).

## 1. Introduction

It is known that the pseudo 3D visualization is the generation of 3D stereo frames based on ready-made single 2D images (image based 3D synthesis) [1-3]. The used methods and algorithms are based on the original images analysis. At the same time, creating their depth maps with that, helps the 3D restoration and replaces the transformation of the original image into a stereo pair [1, 2, 4-11].

In practice, to create a high quality of pseudo 3D stereo, the technologies with “manual” source frames marking are used to create depth maps [1, 2, 4-6]. However, this makes it practically impossible to use them for autonomous online 3D pseudo-stereo generation. In [1, 2], authors reviewed examples of well-known industrial practical implementations of pseudo 3D synthesis in modern 3D TVs/displays. They also stated that the description of their implementations are closed and have some disadvantages like g-hosting, insufficient images spatial depth, etc. In addition, it is indicated that such systems usually use mandatory hardware support for specialized processors. All this confirms the need for improved methods and principles for accelerating the 3D pseudo-stereo synthesis.

In [1, 2], the authors propose a general strategy and modified algorithms for the 3D

pseudo-stereo synthesis. A practical performance study of the proposed solutions and the 3D pseudo-stereo images synthesis characteristics, with/without Compute Unified Device Architect (CUDA) is also carried out.

This article is a continuation of the authors' previous research [1-3] and it is considered as a part of a new computer-aided synthesis system for 3D pseudo-stereo images/video based on hardware support General Purpose Graphics Processing Unit (GPGPU) that allows accelerating the synthesis process. We discuss the CUDA implementation of the main stages of the 3D pseudo-stereo synthesis and the 3D pseudo-stereo characteristics image/video synthesis in Ultra-High Definition (UHD) 4K resolution with a GPU (support for the GPU).

The main contributions of this work are summarized as follows:

- *Using GPGPU to DIBR realization:* Based on GPGPU technology and using modern GPUs, we perform an implementation and a practical study of image based 3D synthesis using the algorithms and approaches proposed in [1, 2].
- *Experimental study and validation:* We implement the characteristics of generating 3D pseudo-stereo images and video in 4K-UHD resolution on modern GPUs with a large number of CUDA-cores. A comparative study is also conducted using the predictions previously made in [1, 2].

Apart from those tow contributions, we identify the main directions for further research on 3D pseudo-stereo generation on high-resolution content, in order to improve the efficiency and the process speed.

The reminder of this paper is organized as follows: Section 2 discusses the related work. Section 3 presents the conversion process of images. In section 4, we show how CUDA is applied for converging images to 3D Pseudo-Stereo. In section 5, we provide the experimental results. Finally, Section 6 is devoted to the conclusion and the future work.

## 2. Related Works

Several methods and algorithms that have been used for “image based 3D synthesis” are based on the original content analysis (images, video) and the creation of their depth maps. This is the basis for Depth Image Based Rendering (DIBR) “pseudo-3D conversion”, transformation of the original image into a stereo pair [1-11].

In order to obtain a “high-quality” pseudo 3D-stereo, technology is applied in practice with manual or semi-automatic marking of source frames to create depth maps and DIBR [1, 2, 4-6, 10]. This makes it impossible to autonomously (without frame marking by the operator), ensure the online generation of high-quality 3D Pseudo-Stereo, especially in the video case.

In [1-3] well-known industrial practical implementations examples of pseudo 3D synthesis in modern 3D monitors are considered. These implementations description are closed. They are characterized by inadequate quality of the received content g-hosting, poor spatial depth of images, etc. It is shown that such systems usually use mandatory hardware support for specialized processors.

In [4-7] still indicated that DIBR is a key technology in image-based 2D-to-3D conversion. Besides this method uses different algorithms that give different quality results. Additionally, it indicates that industrial implementations 2D-to-pseudo-3D conversion and several real-time implementations incorporated into 3D-TVs, sold as stand-alone equipment (stereoscopic image processor), or incorporated into software packages are generally “know-how”. However, the quality of the resulting stereoscopic images, and the depth sensation, is still an outstanding issue that requires more research. In [4], it is argued that the best way for image based 2D-to-3D conversion is to use a specialized real-time 3D rendering processor for 2D-to-3D conversion on stereoscopic displays based on a block-based 2D-to-3D Conversion with Bilateral Filter. Geometry based 2D-to-3D conversion technology and the practical implementation of a system with semi-automatic frame marking, that gives a good visual quality of the synthesized images, is considered in [5]. In [6], Depth methods are considered as the basis for DIBR. It is indicated that for an effective solution of Depth Image Based 2D-to-3D Conversion problems, a Geometry-based approach and the use of Cross Bilateral Filtering is preferred. In [7], a 2D-to-3D conversion system using edge information is considered, that is taken as the basis of the authors' studies [1, 2] in view for acceleration through parallelization.

In [8], the authors have proposed an approach for the disparity mapping of stereoscopic 3D videos. For that, they adopted a nonlinear method that takes advantage of a set of disparity mapping operators. The authors followed a new strategy that, based on stereoscopic warping, it consists of computing disparity and image based saliency estimates.

In [9], a Stereo-to-Multiview video conversion approach has been proposed in the context of glasses-free Multiview displays. In this approach, the authors defined a two-step mapping algorithm in order to preserve the scene’s artistic composition and perceived depth. This algorithm, which is based on a non-linear function, compresses and enhances the scene depth to the depth range of an autostereoscopic display. Then, to return the multiview output, the authors defined an adapted image domain-warping algorithm.

To generate the depth map in an automatic fashion, the authors in [10] employed boundary information in their 2D-to-3D video conversion method. Gaussian model and super-pixel algorithm are also exploited in the detection of foreground objects and edge information. Other algorithms are also used to improve or update the resulting edge information, such as Sobel edge detection and the thinning algorithm. Finally, to produce the 3D video, the authors used depth image-based rendering (DIBR).

In [11], the authors have dealt with disparity adjustment, while taking into account the comfort constraint. In their proposed framework, they started by analyzing two attributes of stereoscopic 3D video (motion and disparity). After that, a discomfort metric was derived based on comfort models. A disparity manipulation step is, finally, applied using a warping strategy in order to generate the output video.

Based on the above discussions, the real-time solution of DIBR problems is urgent and there is a need to improve the performance and quality of Depth Image Based 2D-to-3D Conversion, especially for high-resolution images and video.

Table 1 presents a clear view on the above-discussed works.

It can be seen from Table 1 that the majority of works have either addressed the automatic 2D to 3D conversion issue [5, 6, 7, 9, 10] or focused on disparity manipulation [8, 11] (e.g., mapping, adjustment, marking etc.). Although most of them concentrated on 2D-3D data type [6, 7, 10], others [5, 8, 9, 11] have oriented their efforts on specific video types such as 3D-TV. As for the conversion constraints, we noticed various ones like motion, comfort, sensitivity, spatial frequency, range, sensitivity, gradient and velocity information, etc. To meet their goals, researchers have used a panoply of techniques.

We give as examples, geometric scene categorization, image partitioning, bilateral filtering, nonlinear and stereoscopic warping, Gaussian models, Sobel detection, etc.

## 3. Convert Images to 3d Pseudo-Stereo

In [1, 2], the authors proposed and considered in detail the general strategy for 3D pseudo-stereo image synthesis based on [6, 7] and modified algorithms for its implementation.

Due to the marked closeness of the majority of existing technologies of "intelligent" creation of pseudo 3D that are based on the analysis of literary sources [1-11], the process of obtaining 3D pseudo-stereo from 2D is proposed to be implemented following the scheme in Fig. 1.

The above scheme consists of six core components, which allow converting an input image to a standard 3D stereo image:

- *Depth Map:* represents the process for generating Depth Map as a basis for DIBR.
- *Filtered Depth Map:* depicts the cross bilateral filtering process for "smooth" depth distribution on the map, and therefore improving the quality of DIBR.
- *Left Frame:* is the process for generating the Left Frame of 3D pseudo-stereo Image, based on the original Image and Filtered Depth Map.
- *Right Frame:* as in the Left Frame, this process is responsible for generating the right frame.
- *Post-Processed Left Frame:* this process allows improving the quality of the Left Frame (removing "holes" in the image) based on the Inpainting algorithm.
- *Post-Processed Right Frame:* is dedicated to the improvement of the Right Frame's quality, by exploiting the Inpainting algorithm.

Our method is based on the 2D-to-3D conversion system using edge information [6]. It gives good practical quality results, and the block image processing mechanism potentially permits acceleration through the parallelization. The process main stages and their feats are described in [1, 2]. Thus, a generalized method for obtaining a 3D stereo frame from an image can be represented as in Fig. 2.

The main direction of the acceleration of such systems for the 3D pseudo-stereo image synthesis is the implementation of the process stages on a parallel computer system using graphics processors (GPUs). It is also very important to study the possibility of generating such images in the recently standard 4K format in a real time. It should be noted that in our previous works, we did not consider a practical study of pseudo 3D-stereo synthesis based on GPGPU technology on modern GPUs with a large number of CUDA-cores, to assess the possibility of carrying out the real-time process high-resolution's content conversion.

In this paper, the practical implementation of image-based 3D pseudo-stereo conversion on modern GPUs and obtaining its characteristics for 4K resolution content has a practical value. It also allows identifying the future directions towards increasing the efficiency and the speed of 3D pseudo-stereo content generation.

## 4. System Implementation for Converting Images to 3d Pseudo-Stereo Using Gpu and Nvidia Cuda

Functionally, a system for converting images to 3D pseudo-stereo using GPU and NVidia CUDA technology will not differ from the system described in section 1. Figure 1 and the proposed method for the 3D pseudo-stereo frame synthesis can depict the overall workflow, as well.

As can be seen from the proposed method, each transformation stage is a continuation of the previous stage. The algorithm is sequential as it is impossible to perform simultaneously several stages of processing related to the current image or video sequence frame. Each process stage is analyzed for the possibility of its implementation on the GPU-CUDA architect. [1, 2, 6, 7], in accordance with the NVidia recommendations, parallelization of non-iterative and non-recursive algorithms. Since the algorithm's results directly depend on the previous step's results. That makes the program virtually linear and does not improve the performance [1-3, 12]. From all the obtained stages: A 3D pseudo-stereo image (the proposed method), the generating stages Depth map and Inpainting, it should be noted that the generation algorithm does not give significant performance gains and is associated with a large transfer amounts of high-resolution image data between CPU and GPU memories [1, 2, 6, 7].

Even with implementing the Inpainting stage, it is still poorly suited for the implementation on GPU-CUDA. A simplified solution is proposed in [1, 2] to allow the implementation of this stage on GPU-CUDA with little practical loss of processing quality. The remaining stages of the image processing are not strictly sequential, iterative, recursive and be implemented on GPU-CUDA. Seven steps must be performed in order to convert an image to a 3D pseudo-stereo Image:

- *CPU Map generation:* Depth Map Generation for performing DIBR on the CPU, being that the generation algorithm is iterative and does not give significant performance gains on GPU realization.
- *GPU Cross Bilateral Filtering:* represents the GPU Depth Map filtering organization based on Cross-bilateral filtering to improve the quality of DIBR.
- *GPU Left/Right Frame reconstruction:* consists on the Left Frame (respectively Right Frame) GPU reconstruction of 3D pseudo-stereo Image based on the original Image and Filtered Depth Map.
- *GPU Left/Right Frame Inpainting:* is responsible for improving the quality of the Left Frame (respectively Right Frame) based on the Inpainting algorithm on the GPU.
- *GPU-Stereo 3D Pair formatting (post processing):* forming of standard 3D stereo frame on GPU.

For the experiments, the implementation of process stages on CUDA is carried out. A CUDA code fragment example for implementing the Inpainting algorithm [1, 2] is shown in Fig. 3.

**Table 1. Approaches comparison.**

| Approach | Goal | Data type | Constraints | Adopted technique |
|---|---|---|---|---|
| **[5]** | Semi-Automatic video conversion | 3D-TV | Geometric scene information | Geometry based conversion with semi-automatic frame marking |
| **[6]** | Depth map estimation | 2D-3D | Geometric scene information | Geometric Scene, Categorization image partitioning |
| **[7]** | Automatic conversion | 2D-3D | Edge Information | Bilateral filtering |
| **[8]** | Disparity mapping | Stereoscopic 3D video | Range, sensitivity, gradient and velocity information | Nonlinear and stereoscopic warping |
| **[9]** | Conversion Optimization | Glasses-free and Multiview displays | spatial frequency and disparity information | Nonlinear function and two-step mapping |
| **[10]** | Video conversion | 2D and 3D | Boundary information | Gaussian model, Sobel detection and DIBR |
| **[11]** | Disparity adjustment | S3D | Comfort/Discomfort Disparity and motion information | Disparity mapping, Warping based manipulation |
| **Our approach** | Automatic conversion | 2D-3D | Regions Boundary information | Simplified technique with Bilateral filtering, real time GPGPU implementation |

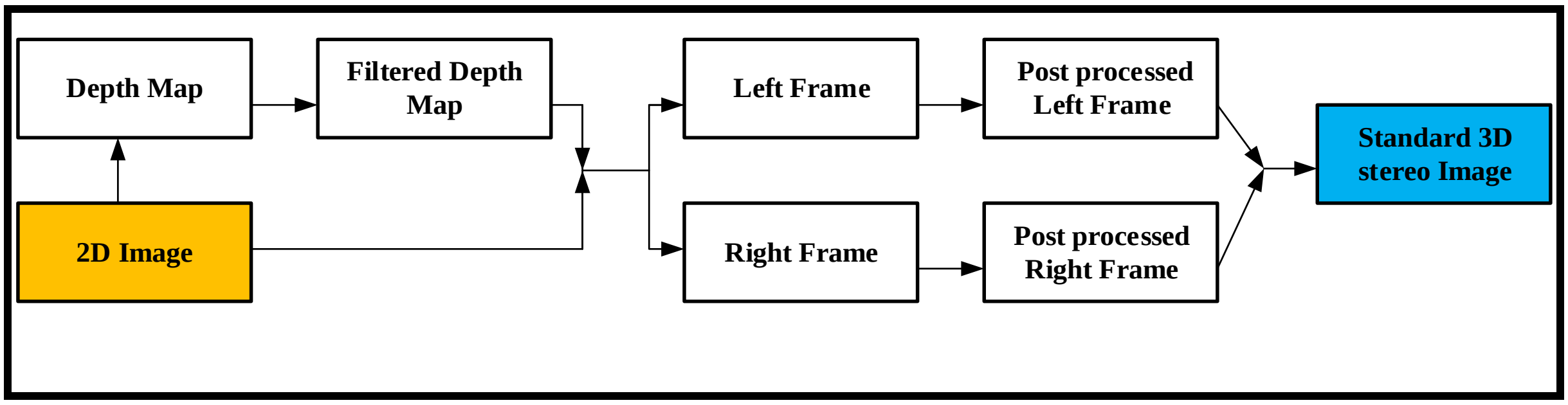

**Fig. 1. Proposed 3D pseudo-stereo conversion scheme.**

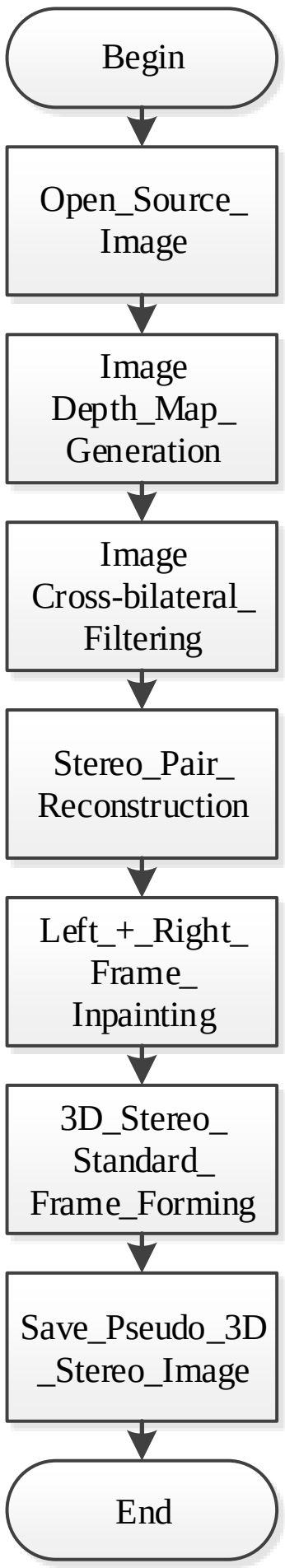

**Fig. 2. 3D pseudo-stereo Synthesis architect.**

| **Inpainting Algorithm** | **Inpainting on CUDA (fragment)** |
|---|---|
| **Begin**<br>Create Frame_Blocks<br>**Do** (*k*=0 **to** *k*= Frame_Blocks)<br>**Begin**<br>Initialization of pixel damage list_k ();<br>Initialization of matrix of all image pixels ();<br>**Do** (*i*=0 **to***i*=Frame_Block_width)<br>**Begin**<br>**Do** (*j*=0 **to***j*=Frame_Block_height)<br>**Begin**<br>**If** current _pixel (*i*,*j*) is damaged<br>**Then**<br>**Begin**<br>Add current_pixel (*i*,*j*) to the list_k of_damaged_pixels;<br>**End**<br>**End**<br>Number_of_damaged_pixels(k):= Number_of _elements In list_k of_ damaged_pixels;<br>**While** number of _damaged_ pixels(k)!=0<br>**Begin**<br>**Do** (i=0 to i=Number_of_elements_list_k_of_damaged_pixels)<br>**Begin**<br>**If** current_pixel (i,j) is damaged<br>**Then**<br>**Begin**<br>Get_number_of_damaged_neighbors (current_pixel (i,j));<br>**If** number_of_undamaged_neigbours >=2<br>**Then**<br>**Begin**<br>Change current_pixel(i,j) by NOT_damaged;<br>Number_of_damaged_pixels(k)--;<br>Compute_color_of_current_pixel(i,j);<br>Set_value_of_current_pixel_of_result_representation(i,j,Pixel_color(i,j));<br>**End** | int row = thread.blockIdx.y*thread.blockDim.y + thread.threadIdx.y;<br>int col = thread.blockIdx.x * thread.blockDim.x + thread.threadIdx.x;<br>int size = ImWidth * ImHeight;<br>…<br>int xLeft = row;<br>int xRight = row;<br>if (FiltDepthVal[col * ImWidth + row] > 150) {<br>xLeft = (int)(row - Base/2*FiltDepthVal[col*ImWidth+ row]/255);<br>xRight=(int)(row+Base/2*FiltDepthVal[col*ImWidth+row]/ 255);}<br>else {<br>xLeft=(int)(row+Base/2*(1-FiltDepthVal[col*ImWidth+row]/255));<br>xRight=(int)(row-Base/2*(1-FiltDepthVal[col*ImWidth+row]/255));}<br>if ((xLeft >= 0) && (xLeft < ImWidth)) {<br>resR[col * ImWidth + row] = srcR[col * ImWidth + xLeft];<br>leftR[col * ImWidth + row] = srcR[col * ImWidth + xLeft];<br>leftG[col * ImWidth + row] = srcG[col * ImWidth + xLeft];<br>leftB[col * ImWidth + row] = srcB[col * ImWidth + xLeft]; }<br>else {<br>resR[col * ImWidth + row] = srcR[col * ImWidth + row];<br>leftR[col * ImWidth + row] = srcR[col * ImWidth + row];<br>leftG[col * ImWidth + row] = srcG[col * ImWidth + row];<br>leftB[col * ImWidth + row] = srcB[col * ImWidth + row];}<br>if ((xRight >= 0) && (xRight < ImWidth)) {<br>resG[col * ImWidth + row] = srcG[col * ImWidth + xRight];<br>resB[col * ImWidth + row] = srcB[col * ImWidth + xRight];<br>rightR[col * ImWidth + row] = srcR[col * ImWidth + xRight]; |
| **End**<br>**End**<br>**End**<br>**End**<br>**End**<br>**End**<br>**End**<br>**End** | rightG[col * ImWidth + row] = srcG[col * ImWidth + xRight];<br>rightB[col * ImWidth + row] = srcB[col * ImWidth + xRight]; }<br>else {<br>resG[col * ImWidth + row] = srcG[col * ImWidth + row];<br>resB[col * ImWidth + row] = srcB[col * ImWidth + row];<br>rightR[col * ImWidth + row] = srcR[col * ImWidth + row];<br>rightG[col * ImWidth + row] = srcG[col * ImWidth + row];<br>rightB[col * ImWidth + row] = srcB[col * ImWidth + row];<br>} |

**Fig. 3. Inpainting algorithm and its CUDA implementation.**

Note that the optimizing possibility on CUDA in the given stages implementation is not considered in the present work and will be the subject of further research.

## 5. 2d-to-3d Pseudo-Stereo Conversion Experimental Study for 4k Uhd

In [1, 2], an experimental study is proposed. Algorithms were carried out for converting images to 3D pseudo-stereo for original images frame formats that do not exceed HD (1920 x 1080) on available GPUs and with a small number of CUDA cores. An assumption is made about the real-time processing possibility of UHD image frames (3840 x 2160) on modern powerful GPUs with more than 2000 cores. The research goal is to verify the made assumptions and to study the practical implementation results of 3D pseudo-stereo synthesis for 4K frames on modern available GPUs.

For the tested configuration, we have implemented our solution on a system with Core™2 Quad Q9550 2.83 GHz/4 GB, while considering four GPU types [13, 14]:

- NVidia GeForce GTX 260 (192 CUDA Core)
- NVidia GeForce GTX 1050 (640 CUDA Core)
- NVidia GeForce GTX 760 (1152 CUDA Core)
- NVidia GeForce GTX 1070i (2432 CUDA Core)

Note that the GeForce GTX 260 is used also in the previous experiments, which are described in [1, 2], while the GeForce GTX 760 and GTX 10x [13, 14] are working in the backward compatible PCI-e 2.0 x16 interface due mode, in order to be applied to the test bench motherboard feat’s. That somewhat limited their overall maximum performance in the system. The studies are carried out for images obtained from an Olympus SP-720UZ portable digital camera [15], in 4K resolution and on a UHD 4K demo video from Sony “Sony Mexico UHD 4K Demos” [16]. The 3D pseudo-stereo experimental generation results for 4K images are presented in Fig. 4 and Fig. 5.

Figure 6 shows the experimental results of computation time (in seconds for anaglyph and anamorphic test image formats) related to the test bench GPUs. Experiments are carried out without considering:

- The time of generating depth map.
- The disk access time for reading and writing the original and resulting images.
- The CUDA code optimization.

That is, it is estimated "p" time of non-optimized computations, where "p" denotes the "pure" computation, those excluding disk and other operations.

Based on the obtained experimental data results, a comparison is made with the forecast data results presented in [1, 2], as shown in Fig. 7.

Experimental data are almost consistent with the previous performed forecasts [1, 2]. Some differences in the obtained experimental and forecasted data are explained as follow:

- The progressive changes in the modern NVidia GPUs micro-architect.
- The operation increased frequency of their cores.
- Memory and interfaces [13, 14], that are not considered when making forecasts [1, 2].

The obtained data confirm the computing 4K 3D pseudo-stereo images possibility using modern GPUs in real time. In addition, experiments were conducted on the synthesis 4K 3D-pseudo-stereo video on UHD 4K demo video "Sony Mexico UHD 4K Demos" [16]. The experimental results on a test stand are presented in Table 2 and Fig. 8.

The video processing time depends on the video resolution, on the number of frames per second and the video duration, as well as on the information in the image itself. The experimental results show that most of the conversion time for the GPU implementation is not spent on the video frame processing, but rather on the constantly ongoing data exchange process between the video card, the processor and the RAM. As well as the video encoding process with a standard video codec and disk, read/write video streams operations.

As studies, result of 4K content pseudo-stereo synthesis on test GPUs, confirmation and consistency with those previously made by the authors is obtained. They show the possibility of real-time 3D pseudo-stereo restoration by the proposed tools using the support of modern Graphics Processing Unit/Graphics Processing Clusters (GPU/GPC). The average practical acceleration of the 3D pseudo-stereo synthesis process on a test stand using the NVidia GeForce GTX 1050 GPU was about 60.6 and 6.9 times for images and videos, respectively.

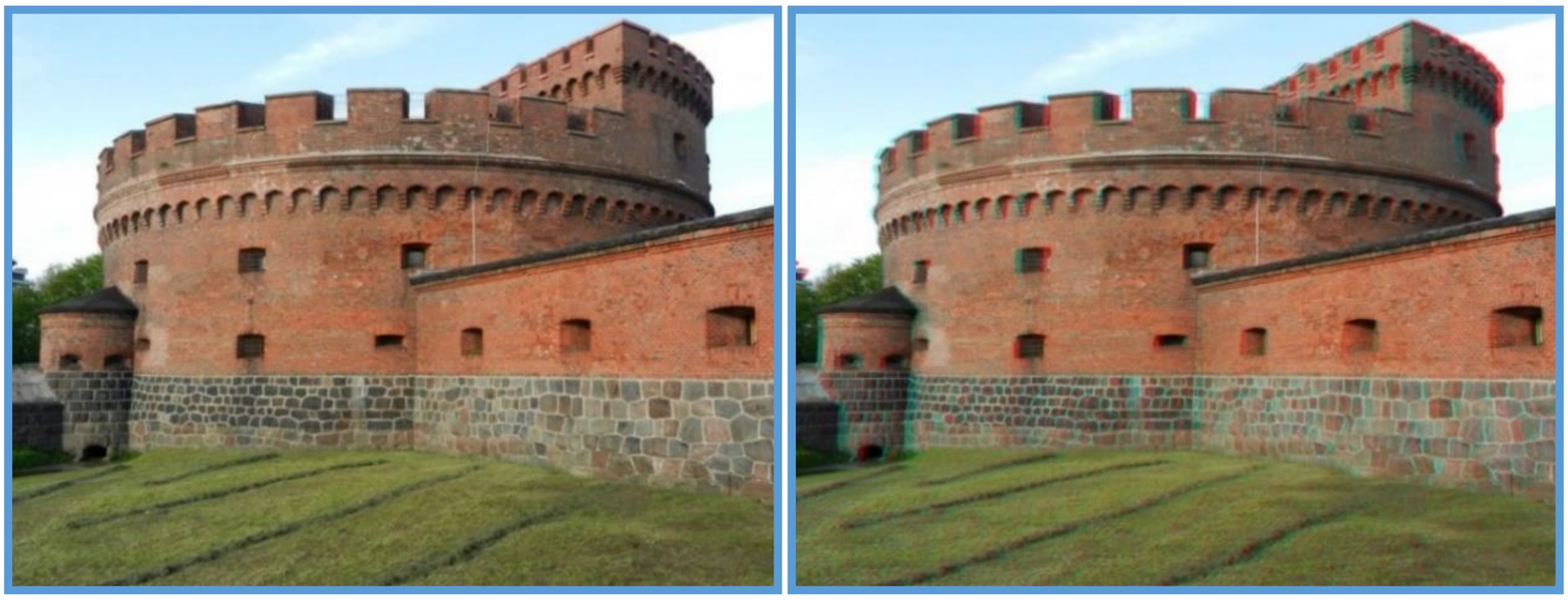

**a) Original image** **b) Stereo anaglyph**

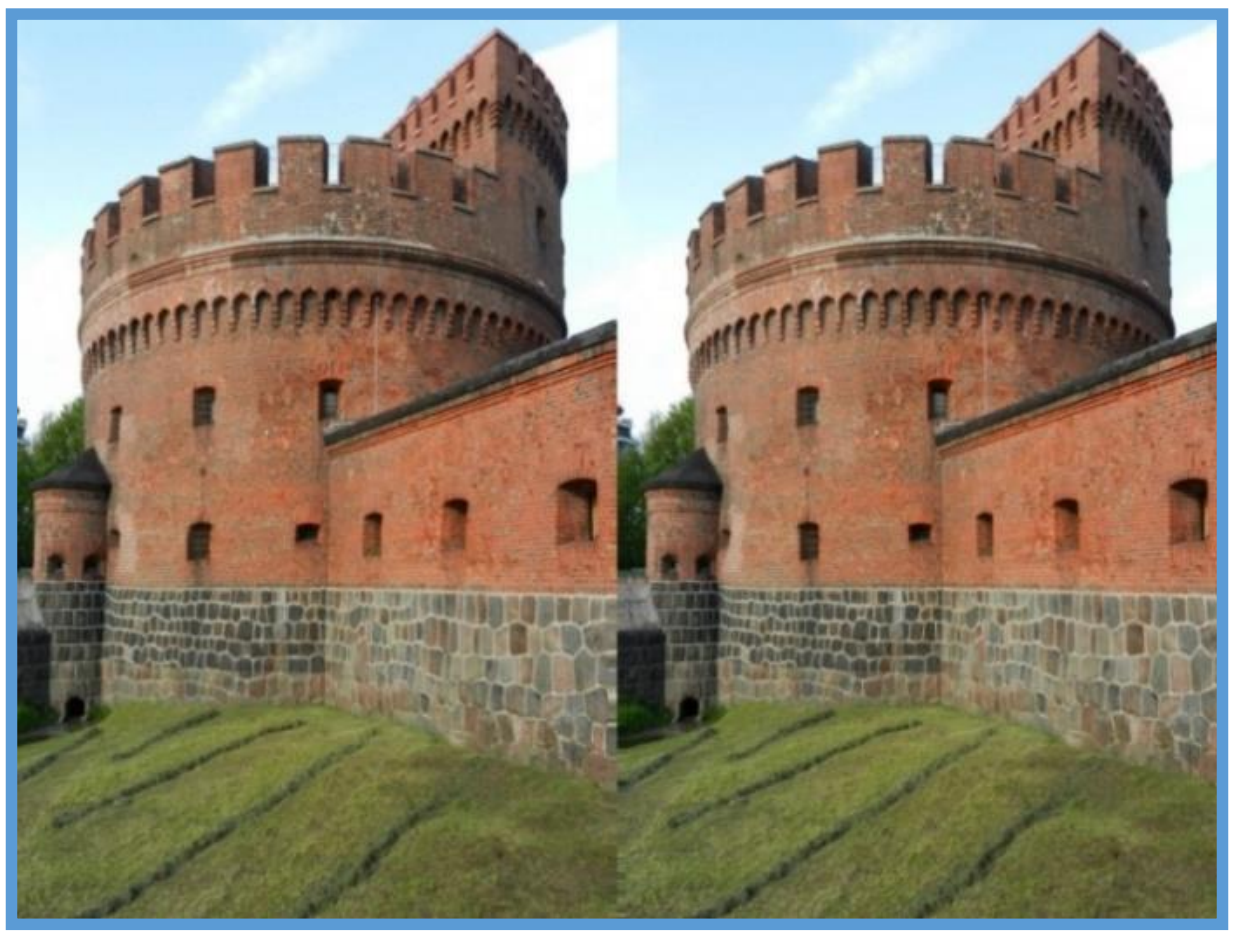

c) Anamorphic HSBS stereo pair

Fig. 4. 4K-3D-pseudo-stereo image synthesis results for image 1.

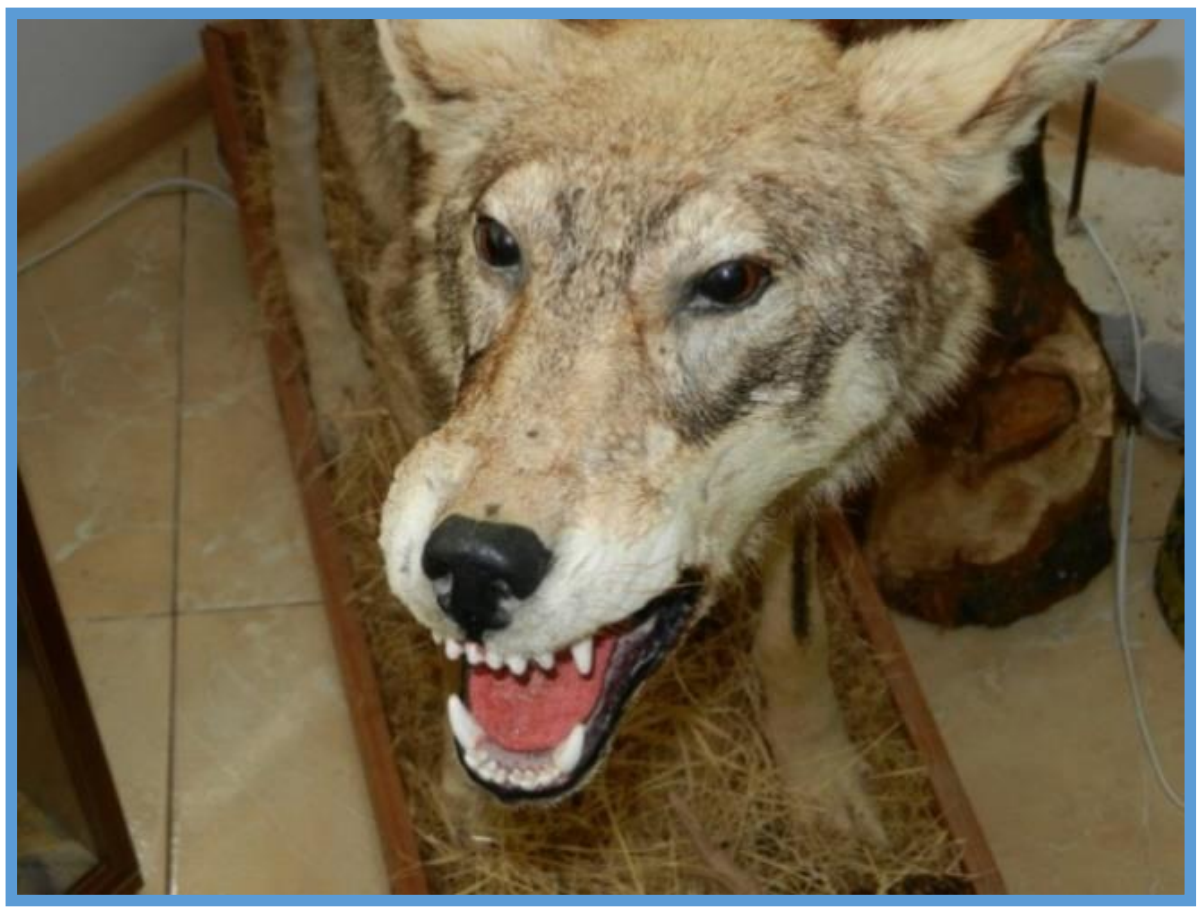

a) Original image

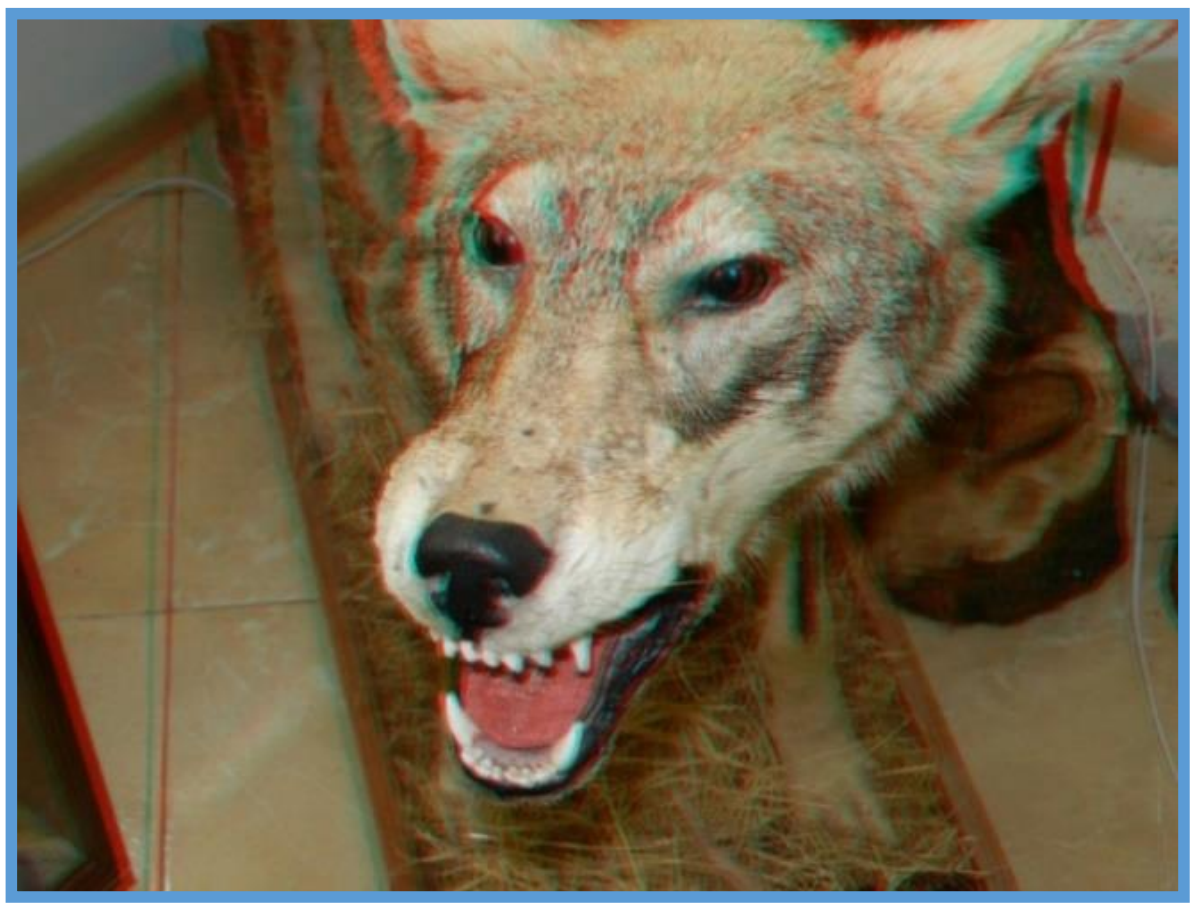

b) Stereo anaglyph

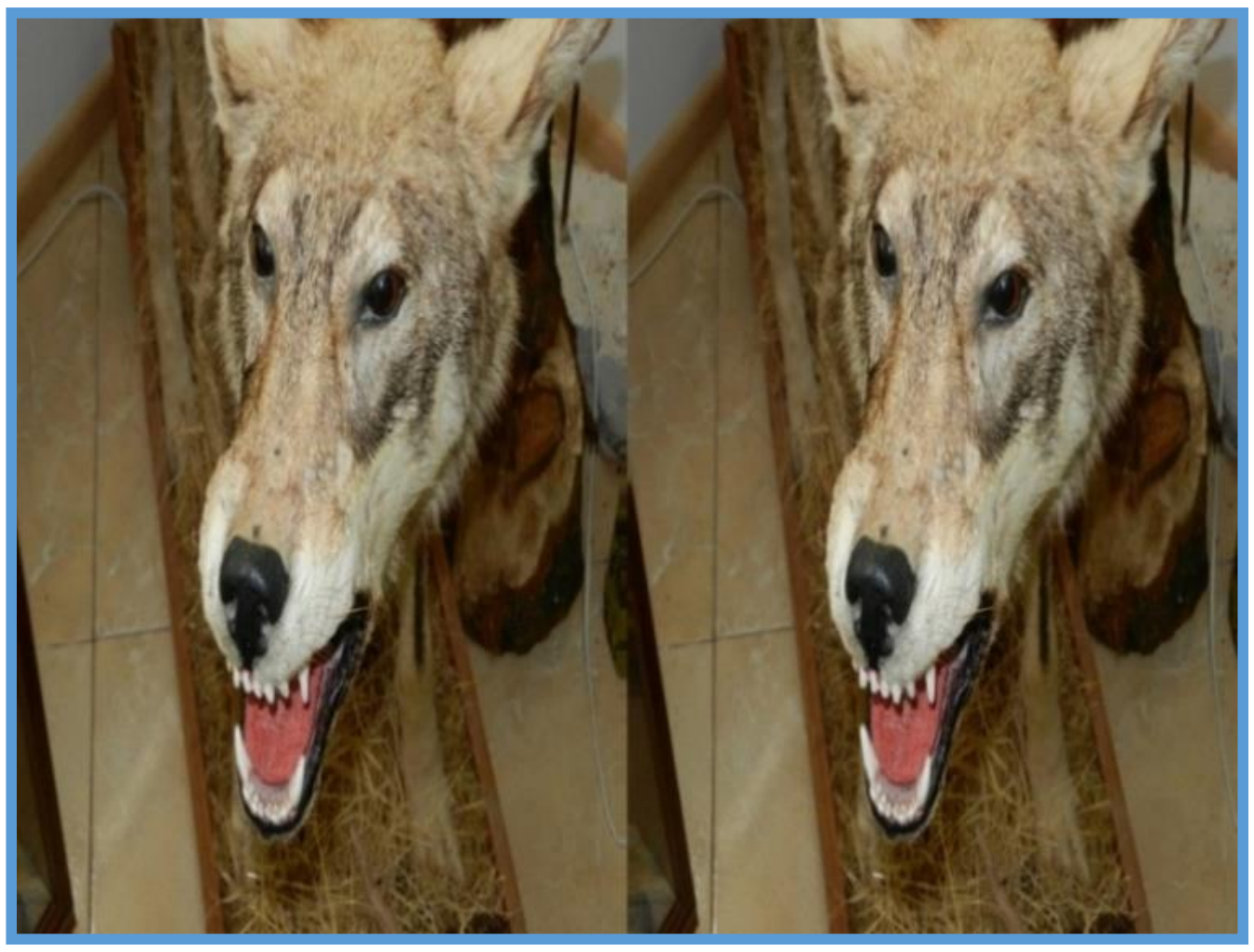

c) Anamorphic HSBS stereo pair

Fig. 5. 4K-3D-pseudo-stereo image synthesis results for image 2.

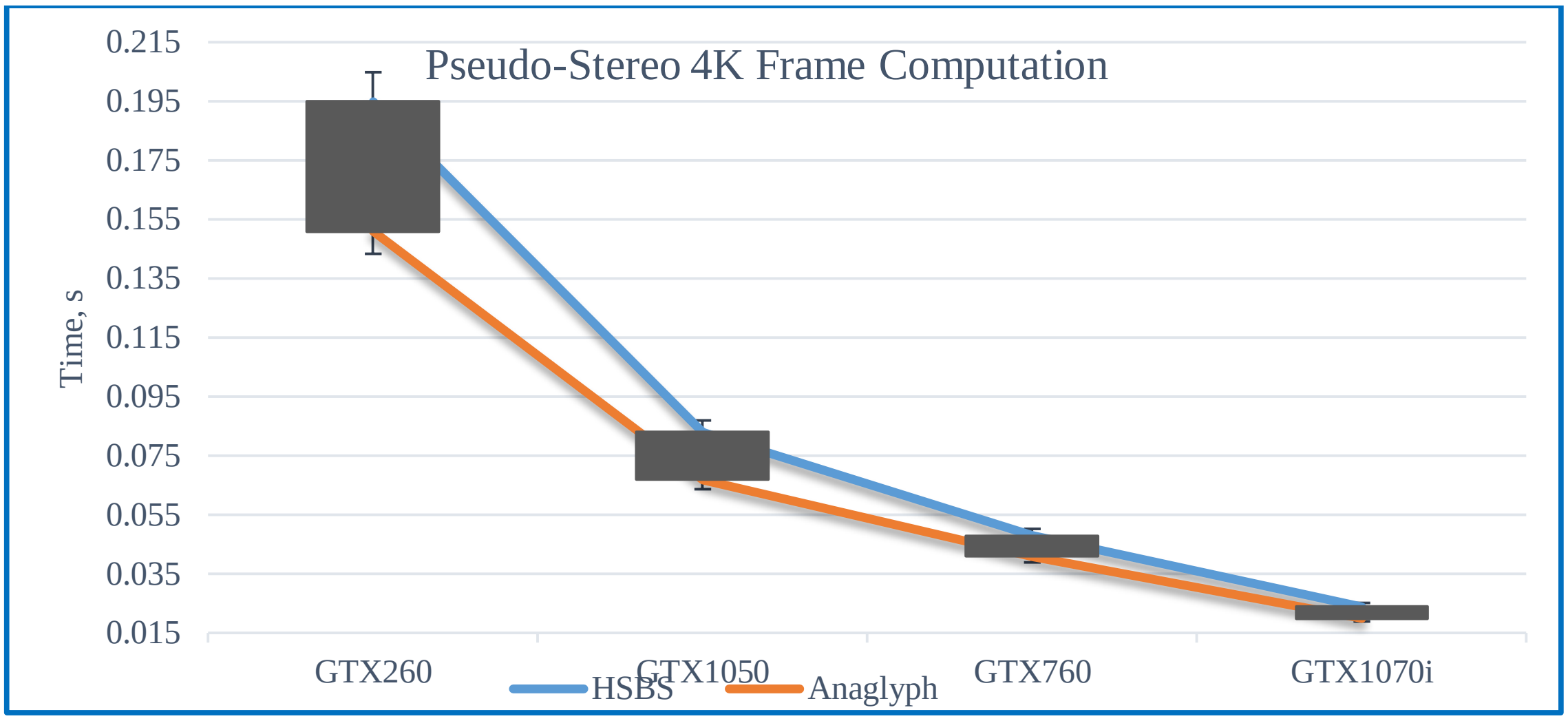

**Fig. 6. 4K pseudo-stereo image computation time.**

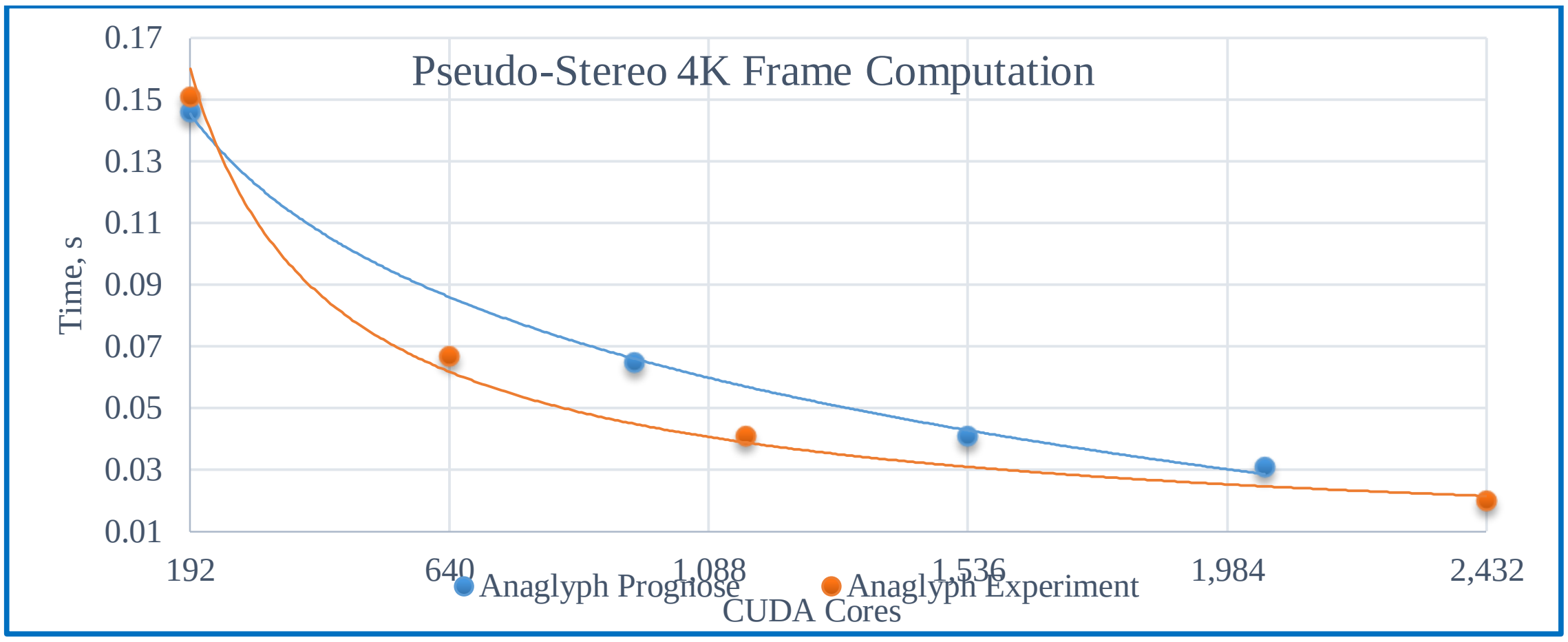

**Fig. 7. Experimental comparison and prediction data for the 4K pseudo-stereo images synthesis.**

**Table 2. 3D Pseudo-Stereo generation of UHD 4K video.**

| Test video | Converting video duration to 3D pseudo-stereo (anaglyph) | | Acceleration |
|---|---|---|---|
| | without GPU | with GPU NVidia GeForce GTX 1050 | |
| 4K UHD 3840x2160,<br>60 fps, 1 min 55 s | 147 min 25 s | 23 min 9 s | 6.89 times |

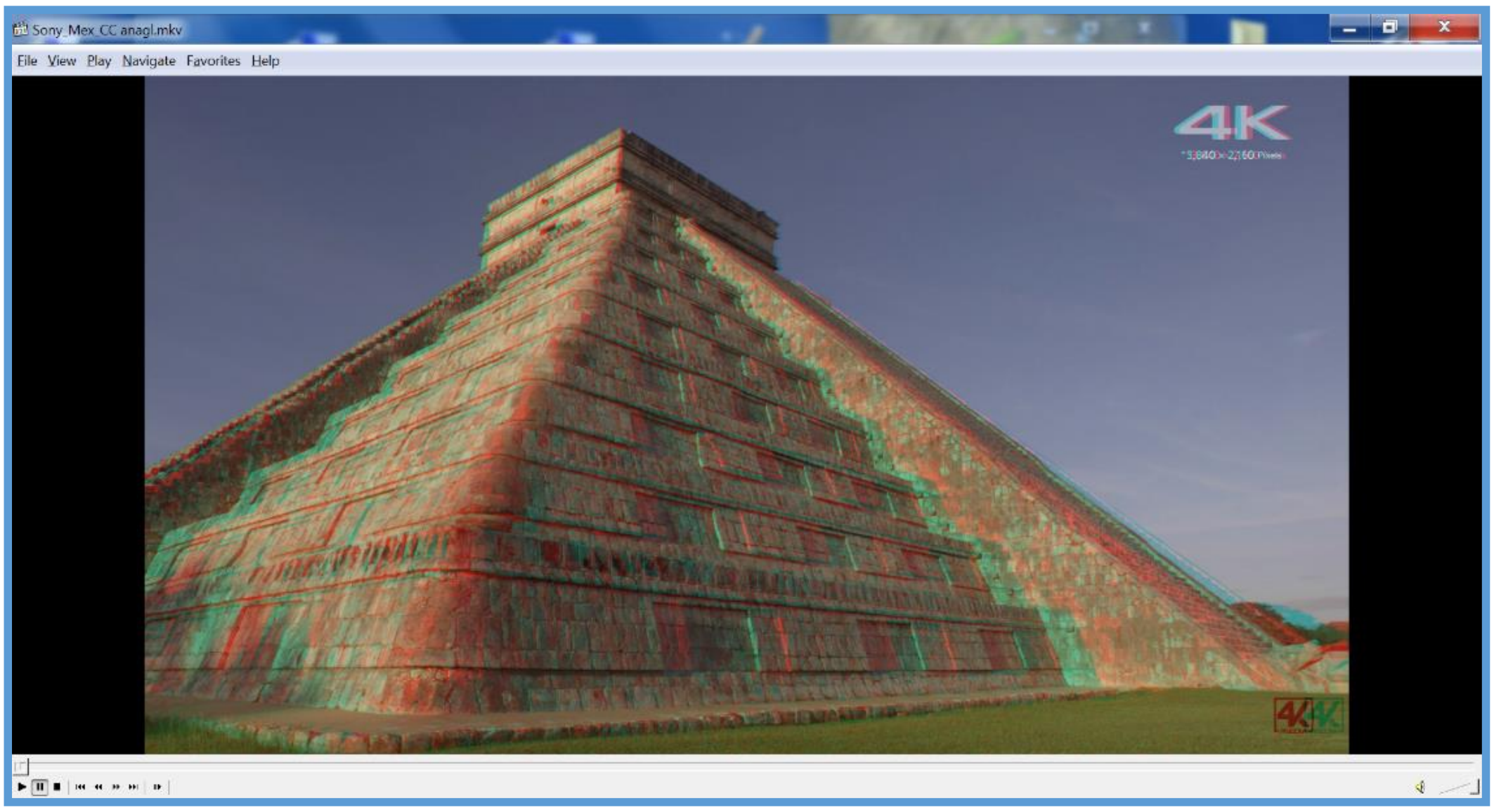

**Fig. 8. UHD 4K 3D pseudo-stereo video generation.**

## 6. Conclusions

By considering, the general strategy of 3D pseudo-stereo synthesis proposed by the authors, this article conducted a practical study of the 3D pseudo-stereo synthesis images and video using hardware and software support for process acceleration (based on GPGPU technology).

The non-optimized CUDA implementation of the 3D pseudo-stereo synthesis main stages has been considered. The practical implementation study and computations characteristics on modern GPUs for synthesis images and video in UHD 4K resolution has been performed. It was proven the possibility of real-time 3D pseudo-stereo restoration using the hardware-software support of modern GPU/GPC. The average practical acceleration of the 3D pseudo-stereo synthesis was satisfactory for both images and videos.

Some differences in the obtained experimental and forecast data are explained by the progressive changes in the modern NVidia GPUs micro-architect, such as an increase of the cores, memory and data interfaces frequency, which was not previously taken into account when making forecasts.

The CUDA code optimization and the configuration optimization of the CUDA computer network will provide additional acceleration of the target process.

Finally, with the goal to increase the speed and efficiency of the 3D pseudo-stereo content synthesis process, we intend to analyze and implement the modifying possibility of Cross-bilateral Filtering at the stage of Depth Map Generation, based on GPGPU technology.

**Conflict of Interest:** The authors declare that they have no conflicts of interest.

# تصور استريو ثلاثي الأبعاد غير حقيقي مع دعم وحدة معالجة الرسومات للأغراض العامة

**أنس محمد العريقات[١] و سيرجي أناتوليفيتش زوري[٢]**

*[١] قسم الأمن السيبراني، كليات عنيزة، كلية الهندسة وتقنية المعلومات، عنيزة، ص.ب ٥٣٧١، المملكة العربية السعودية، و[٢] قسم علوم وتقنيات الكمبيوتر، جامعة دونيتسك التقنية الوطنية، دونيتسكا، ص.ب ٨٥٣٠٠، أوبلاست، أوكرانيا*

anas_oraiqat@hotmail.com

*المستخلص*. تناقش هذه المقالة دراسة نظام الكمبيوتر لإنشاء صور ومقاطع فيديو ثلاثية الأبعاد مجسمة غير حقيقية باستخدام دعم الأجهزة والبرامج لتسريع عملية التوليف بناءً على تقنية وحدة معالجة الرسومات للأغراض العامة. استنادًا إلى الاستراتيجية العامة للتركيب المجسم ثلاثي الأبعاد غير الحقيقي الذي اقترحه المؤلفون سابقًا، تأخذ طريقة تنفيذ حساب موحد لمعامرية الجهاز مع الأخذ بالاعتبار مراحل التنفيذ الرئيسية للتركيب المجسم ثلاثي الأبعاد غير الحقيقي: (١) دراسة التنفيذ العملي؛ (٢) خصائص التوليف للحصول على الصور؛ (٣) الفيديو بدقة 4K فائقة الوضوح باستخدام وحدة معالجة الرسومات. على التوالي مع هذه النتائج لاختبار محتوى 4K على أنظمة التقييم باستخدام وحدة معالجة الرسومات، يتم الحصول على متوسط تسريع يبلغ ٦٠,٦ و٦,٩ مرة للصور ومقاطع الفيديو. تظهر نتائج البحث تناسقًا مع التوقعات المحددة مسبقًا لمعالجة إطارات الصور بدقة 4K. النتائج تؤكد إمكانية توليف خوارزميات ثلاثية الأبعاد مجسمة غير حقيقية في الوقت الفعلي باستخدام دعم قوي لوحدة معالجة الرسومات الحديثة وعناقيد مجموعات معالجة الرسومات.